\documentclass[twocolumn,preprintnumbers,amsmath,amssymbm,prd]{revtex4}
\usepackage{epsfig}
\usepackage{graphicx}
\usepackage{amssymb}

\begin{document}

\title{Comment on: ``Possible Relation between the Cosmological Constant and Standard Model Parameters''}
\author{Shahar Hod}
\address{The Ruppin Academic Center, Emeq Hefer 40250, Israel}
\address{}
\address{The Hadassah Institute, Jerusalem 91010, Israel}
\date{\today}

\begin{abstract}

\ \ \ It has recently been proposed [M. P. Hertzberg and A. Loeb, arXiv:2302.09090] 
that the observed value of the cosmological constant can be related to 
the physical parameters of the Standard Model. 
In the present compact note we point out that the 
derivation of the claimed cosmological-constant-standard-model relation presented 
in \cite{HL} is, unfortunately, erroneous. 
\end{abstract}
\bigskip
\maketitle

An intriguing paper has recently appeared in the physics literature \cite{HL}, in which it is claimed that 
the remarkably small value of the cosmological constant, 
\begin{equation}\label{Eq1}
\Lambda\simeq 10^{-123}l^{-2}_{\text{P}}\  ,
\end{equation}
can be related to the fundamental physical parameters of nature: $G,c,\text{e},m_{\text{e}}$. 
The derivation of the proposed relation presented in \cite{HL} is based on a 
scattering process that involves a charged black hole and a probe which is assumed to take place within 
a Hubble time before the black hole evaporates away.

In the present short note we point out that the derivation of the 
claimed cosmological-constant-standard-model relation presented in \cite{HL} is, 
unfortunately, physically inconsistent. 

In particular, it is assumed in \cite{HL} that the mutual electric interaction between a black hole which 
has the minimally possible non-zero electric charge,
\begin{equation}\label{Eq2}
Q_{\text{BH}}=e\  ,
\end{equation}
and the electron (the particle with the largest charge-to-mass ratio in nature) 
is comparable to, or even stronger than, their mutual gravitational interaction. 

As shown in \cite{HL}, this assumption implies that the mass $M_{\text{BH}}$ of the black hole which is used to derive 
the claimed cosmological-constant-standard-model relation is 
purely specified by the fundamental constants of nature (we use natural units in which $G=c=\hbar=1$ \cite{Noteun}):
\begin{equation}\label{Eq3}
M_{\text{BH}}={{e^2}\over{m_{\text{e}}b}}\  ,
\end{equation}
where $m_{\text{e}}$ is the proper mass of the electron and
\begin{equation}\label{Eq4}
b\gtrsim1\
\end{equation}
is a dimensionless fudge factor introduced in \cite{HL} [it is stated in \cite{HL} that the relation 
$b=O(10)$ is plausible].

We first point out that the charge-to-mass ratio of the black hole which is used in \cite{HL} 
is characterized by the dimensionless relation [see Eqs. (\ref{Eq2}) and (\ref{Eq3})]
\begin{equation}\label{Eq5}
{{Q_{\text{BH}}}\over{M_{\text{BH}}}}={{m_{\text{e}}b}\over{e}}\simeq 4.9\times 10^{-22}b\ll1\  .
\end{equation}
In addition, the black hole used in \cite{HL} is characterized by the dimensionless 
relation [see Eqs. (\ref{Eq1}) and (\ref{Eq3})]
\begin{equation}\label{Eq6}
\Lambda\cdot M^2_{\text{BH}}\sim 10^{-123}\cdot 10^{40}\cdot b^{-2}\ll1\  .
\end{equation}

The characteristic strong inequalities (\ref{Eq5}) and (\ref{Eq6}) imply that the near-horizon region of the 
black-hole spacetime is well approximated by the Schwarzschild relation $g_{tt}\simeq 1-2M_{\text{BH}}/r$. 
In particular, the horizon radius 
of the black hole can be approximated by the Schwarzschild relation
\begin{equation}\label{Eq7}
r_{\text{H}}\simeq 2M_{\text{BH}}\  .
\end{equation}

Taking cognizance of Eqs. (\ref{Eq2}) and (\ref{Eq7}), one finds that the 
electric field strength ${\cal E}_{\text{BH}}=Q_{\text{BH}}/r^2_{\text{H}}$ on the surface 
of the charged black hole is given by the relation
\begin{equation}\label{Eq8}
{\cal E}_{\text{BH}}\simeq{{e}\over{4M^2_{\text{BH}}}}={{m^2_{\text{e}}b^2}\over{4e^3}}\  .
\end{equation}

We now point out the important fact that the analysis presented in \cite{HL} can only be valid if the electric field 
strength (\ref{Eq8}) of the black hole (\ref{Eq3}), which is used to derive 
the claimed cosmological-constant-standard-model relation, is {\it weaker} than the {\it critical} 
electric field \cite{Sch1,Sch2,Sch3}
\begin{equation}\label{Eq9}
{\cal E}_{\text{BH}}\leq {\cal E}_{\text{c}}\equiv {{m^2_{\text{e}}}\over{e}}\
\end{equation}
for quantum production of electron-positron pairs. 

From Eqs. (\ref{Eq8}) and (\ref{Eq9}) we find that the dimensionless inequality
\begin{equation}\label{Eq10}
b\lesssim 2e\simeq{{2}\over{137^{1/2}}}\simeq0.17\
\end{equation}
provides a necessary condition for the validity of the analysis presented in \cite{HL}. 
The analytically derived necessary condition (\ref{Eq10}) contradicts the 
inequality (\ref{Eq4}) that was assumed in \cite{HL}. 
Thus, one is forced to deduce that, unfortunately, the analysis presented in \cite{HL} is erroneous. 

\bigskip
\noindent
{\bf ACKNOWLEDGMENTS}
%\bigskip

This research is supported by the Carmel Science Foundation. I would
like to thank Yael Oren, Arbel M. Ongo, Ayelet B. Lata, and Alona B.
Tea for helpful discussions.


\begin{thebibliography}{99}

\bibitem{HL} M. P. Hertzberg and A. Loeb, arXiv:2302.09090 .

\bibitem{Noteun} One finds the dimensionless relations $e\simeq 1/137^{1/2}$ 
and $m_{\text{e}}\simeq 4.19\cdot 10^{-23}$ in these natural units.
 
\bibitem{Sch1} F. Sauter, Z. Phys. {\bf 69}, 742 (1931); 
W. Heisenberg and H. Euler, Z. Phys. {\bf 98}, 714 (1936); 
J. Schwinger, Phys. Rev. {\bf 82}, 664 (1951).

\bibitem{Sch2} W. T. Zaumen, Nature {\bf 247}, 531 (1974); 
B. Carter, Phys. Rev. Lett. {\bf 33}, 558 (1974); 
G. W. Gibbons, Comm. Math. Phys. {\bf 44}, 245 (1975); 
T. Damour and R. Ruffini, Phys. Rev. Lett. {\bf 35}, 463 (1975).

\bibitem{Sch3} S. Hod, Phys. Rev. D {\bf 59}, 024014 (1999) [arXiv:gr-qc/9906004]; 
S. Hod and T. Piran, Gen. Rel. Grav. {\bf 32}, 2333 (2000) [arXiv:gr-qc/0011003]; 
S. Hod, Phys. Lett. B {\bf 693}, 339 (2010) [arXiv:1009.3695].


\end{thebibliography}
\end{document}